\documentclass{llncs}

\usepackage{graphicx}
\usepackage{tabularx}
\usepackage{hyperref}
\usepackage{graphicx}
\usepackage[british]{babel}
\usepackage{eurosym}
\usepackage{pbox}
\usepackage{url}
\usepackage{xspace}
\makeatletter
\g@addto@macro{\UrlBreaks}{\UrlOrds}
\usepackage{booktabs}
\usepackage[misc]{ifsym}
\usepackage[font=small,labelfont=bf,labelsep=space]{caption}
\begin{document}
\mainmatter

\title{Optional Data Disclosure and the \\Online Privacy Paradox: A UK Perspective}
\titlerunning{Optional Data Disclosure and the Online Privacy Paradox}
 
\author{Meredydd Williams\textsuperscript{(\Letter)}, Jason R. C. Nurse}
\authorrunning{Williams and Nurse}

\institute{Department of Computer Science,\newline University of Oxford, Oxford, UK\\
\email{meredydd.williams@cs.ox.ac.uk}}

\maketitle

\begin{abstract}
Opinion polls suggest that the public value their privacy, with majorities calling for greater control of their data. However, individuals continue to use online services which place their personal information at risk, comprising a Privacy Paradox. Previous work has analysed this phenomenon through after-the-fact comparisons, but not studied disclosure behaviour during questioning. We physically surveyed UK cities to study how the British public regard privacy and how perceptions differ between demographic groups. Through analysis of optional data disclosure, we empirically examined whether those who claim to value their privacy act privately with their own data. We found that both opinions and self-reported actions have little effect on disclosure, with over 99\% of individuals revealing private data needlessly. We show that not only do individuals act contrary to their opinions, they disclose information needlessly even whilst describing themselves as private. We believe our findings encourage further analysis of data disclosure, as a means of studying genuine privacy behaviour.

\end{abstract} 

\section{Introduction}
\label{sec:one}

Privacy principles have been present throughout human history, with the Ancient Greek philosopher Socrates drawing distinctions between the ``outer'' and ``inner'' self \cite{konvitz1966privacy}. Warren and Brandeis' \textit{Right to be Let Alone} \cite{warren1890right} placed privacy within the Western democratic consciousness, and across much of the world it is clearly established as both a legal and a human right \cite{robertson1968united}. However, whilst ephemeral utterances were once lost in the ether, now all our electronic communications are logged, stored and used for later processing. With today's spontaneous conversations persisting decades into the future, privacy is at a crossroads.

Previous studies have suggested individuals care about their privacy. The Pew Research Center \cite{rainie2013anonymity} found that 86\% of surveyed US citizens reported taking steps to remain private online, whilst 88\% in a UK poll claimed to value their privacy \cite{truste2015privacy}. Despite these self-reported surveys, many examples speak to the contrary. Another Pew poll found that 74\% of US respondents used location-based services on their smartphones, allowing their movements to be tracked in real-time \cite{zickuhr2012three}. Employees are dismissed due to embarrassing online disclosures \cite{Mosbergen2015}, burglars survey social networking sites to select their targets, and the use of various technologies leads to an increasing number of privacy risks \cite{Nurse2015xrds}. \cite{bloxham2011most}. Later research \cite{carrascal2013your} found that individuals valued their online browsing history at only \euro 7, the price of a fast-food meal. We are presented with the ``Privacy Paradox'' \cite{barnes2006privacy}, where individuals claim to value privacy but do little to actively protect it.

The novelty of this paper is to examine the paradox empirically, comparing what individuals claim about their privacy actions with their data disclosure behaviour. Previous research \cite{norberg2007privacy} has taken a two-phase approach, judging participants' actions to be less private than their reported intentions. Other work \cite{jensen2005privacy} questioned individuals on their privacy perceptions, before finding they unwisely judged user interfaces. However, these privacy evaluations were distinct from initial questioning, with little research empirically assessing behaviour during replies. Culnan \cite{culnan1995consumer} discovered many demographic factors might influence privacy concern, including age, wealth and education level. However, the majority of previous work has been conducted in the US, with British privacy perceptions rarely considered. In contrast, our research studied how the UK public perceive privacy, how privately do they act, and which privacy-protective technologies do they use. Through this, we examined how different demographic groups view privacy, and whether those who claim to value their privacy act privately with their own data.

In achieving these goals, we first surveyed existing work on the Privacy Paradox. Conventionally, interpretations are split into two groups: opinion-oriented and behaviour-oriented \cite{baek2014solving}. The former considers the disparity due to a lack of user education, whilst the latter judges individuals to exchange privacy for convenience. Whereas Acquisti \cite{acquisti2004privacy} justified user actions through the behavioural economic theory of bounded rationality, Jensen \textit{et al.} \cite{jensen2005privacy} attributed the disparity to individuals' misjudging their own abilities. Next, we built on the position paper by Williams \cite{williams2015study} by conducting a physical survey across several cities in the UK. We analysed our collected data on three levels: firstly, to study how the British public view privacy; secondly, to investigate how different demographic groups regard the topic; and thirdly, to examine whether those who claim to value privacy act privately with their own data. 

The remainder of our paper is structured as follows. Section \ref{sec:two} reviews the literature concerning privacy definitions, demographic studies and the Privacy Paradox. Section \ref{sec:three} then describes our research questions, the methodology of our physical survey, and how we designed against response biases. Section \ref{sec:four} continues by presenting our survey results, performing data analyses, and discussing our findings. Finally, we conclude the paper in Section \ref{sec:five} and consider possibilities for further work.
\newpage

\section{Literature Review}
\label{sec:two}

Before we discuss privacy in greater detail, we should produce a clear definition. Clarke \cite{clarke1999introduction} found distinctions between information privacy, media privacy, interception privacy, and bodily privacy, with Burgoon \cite{burgoon1982privacy} also regarding an informational privacy component important. For the purposes of our work, any reference to privacy concerns information privacy, i.e. ``\textit{the interest an individual has in controlling, or at least significantly influencing, the handling of data about themselves}'' \cite{clarke1999introduction}. This is due to the fact we are studying  data distribution rather than more societal interactions.

Previous research has compared demographic groups based on their privacy opinions. Whilst Han and Maclaurin \cite{han2002consumers} found online privacy concern generally increased with age, other work considered whether younger people might be better at protecting themselves due to their greater knowledge of modern technology \cite{blank2014new}. Sheehan \cite{sheehan1999investigation} found women to be more concerned than men, although other studies \cite{chen2004protecting} have shown that male users tend to falsify their personal data more frequently. As previously mentioned, Culnan \cite{culnan1995consumer} discovered several demographic factors which might influence concern for privacy, including age, wealth and education level. All this previous research was undertaken in the US, whilst our work studies whether demographic factors have an influence in the UK. Furthermore, whilst polls are generally undertaken online, reducing participation from certain demographic groups \cite{evans2005value}, our survey was conducted in an accessible manner on public streets.

The Privacy Paradox has been analysed in a variety of contexts. Barnes \cite{barnes2006privacy} studied the teenage use of social networking sites, finding that whilst teens freely disclose their personal information, they still express outrage when their privacy is invaded. Norberg \textit{et al.} \cite{norberg2007privacy} questioned participants on their willingness to disclose data, before requesting the same information several weeks later through market researchers. They discovered that regardless of the type of information, the disclosure level was far greater than what respondents had initially claimed. Motiwalla \textit{et al.} \cite{motiwalla2014privacy} used an auction scenario to analyse disclosure behaviours, finding stated concerns to be a poor predictor of future actions. Although these studies illuminate the Privacy Paradox, analyses are performed after-the-fact and therefore other variables, such as changes in context \cite{morando2014privacy}, could have an effect. Acquisti \cite{acquisti2004privacy} considered bounded rationality, finding that users might focus on short-term gratification, without considering the long-term privacy risks. Syverson \cite{syverson2003paradoxical} rejected that individuals act irrationally, claiming they weigh costs against benefits in a sensible manner, with this echoed by findings of a January 2016 US poll \cite{rainie2016}. Although economic analyses are enlightening, we decided to investigate the matter empirically to compare opinions with actions. In his comprehensive 2015 literature review, Kokolakis \cite{kokolakis2015privacy} urged future research to make use of data rather than relying on self-reported claims. We combined these approaches, comparing reported opinions with empirical disclosure behaviour.
\newpage

\section{Research Methodology}
\label{sec:three} 

We explored three main research questions (RQ) through our survey on the Privacy Paradox. These are as follows:
\begin{enumerate}
\item How do the UK public regard online privacy, and which privacy-protective technologies do they use?
\item How do demographic groups in the UK differ from each other in their perceptions of privacy?
\item Do those who claim to value their privacy act privately when given the option to disclose sensitive demographic data? 
\end{enumerate}

We undertook a physical survey of the UK's general adult population, choosing London, Birmingham, Cardiff and Oxford for study due to their varied locations and differing population sizes (8.6M, 1M, 350K, and 154K, respectively). We began preparing the survey several months before the study, both through acquiring university ethical approval and to allow for an iterative design process. We also received explicit authorisation from the various city councils to conduct our research. The demographic queries, survey questions and options for reply were updated frequently before the form was finalised. We then undertook a pilot study for one day in Oxford to ensure that the questionnaire was appropriate for our research. The final survey was undertaken in August 2015, with participants asked how they value privacy, how they view their own level of privacy-consciousness, and which tools they use to protect themselves.

We divided the questionnaire into five sections: Required Demographics, Optional Demographics, Opinions, Actions, and Informed Consent. The Required Demographics included gender and age range and were used to study how different sections of UK society regard online privacy. The Optional Demographics allowed empirical analysis of data disclosure and comprised of marital status, employer and Twitter handle. Participants could either choose to answer these questions or select ``Prefer Not To Say'', with only the  presence or absence of this data taken into consideration. The Opinion section asked participants to rate their agreement with privacy statements (for example, ``privacy is of importance to me'') on a five-point Likert scale from ``Strongly Agree'' to ``Strongly Disagree''. We requested these opinions before individuals reported their actions, as a means of reducing priming effects and social desirability bias \cite{fisher1993social}. The final opinion asked participants to self-evaluate their own level of privacy, and this was compared with their reported actions and disclosure behaviour.
 
In the Actions section, participants were questioned on their online habits (for example, ``how often do you share your location on social networking sites?''), with a six-point Likert scale ranging from ``Always'' to ``Never'' with ``Unsure'' and ``N/A'' included. We chose this frequency scale both to reduce the risk of acquiescence bias \cite{churchill1979paradigm}, where participants just agree with the researcher, and to reflect that individuals might protect information differently in different contexts. Both the Opinion and Action questions were designed to be neutral to also reduce this acquiescence risk. The Informed Consent section notified participants of study purpose and redressal procedures, important matters for performing ethical research. We used a chocolate bar reward to incentivise participation, and whilst some individuals might complete the questionnaire just for personal gain, we found most respondents were interested in providing their opinion.

Several risks exist from surveying the general public in a self-reporting manner, and we remained cognisant of possible response biases. Non-response bias \cite{hansen1946problem} could have been encountered if those opposed to disclosing data therefore avoided our survey. Whilst this was considered, we assured participants that their data would be stored securely and that the survey had passed ethical approval procedures. Social desirability bias could influence participants to exaggerate their privacy concern, and therefore we used the amount of disclosed data as a proxy for privacy behaviour. Some respondents might provide further information for completeness or in reaction to demand characteristics \cite{orne1962social}. Furthermore, privacy behaviour is contextualised \cite{morando2014privacy} and individuals might disclose more data knowing that the survey is for academic purposes and conducted through the University of Oxford. However, participants were instructed that these fields were not mandatory and the ``Prefer Not To Say'' options offered a simple alternative. Respondents might claim to be using privacy-protective technologies falsely, either due to misconceptions or to appear privacy-conscious. Jensen \textit{et al.} \cite{jensen2005privacy} found that individuals often claimed to understand a technology, but then were unable to answer simple questions about it. To reduce this risk, we included the fictitious \textit{PrivBrowse} product, in the technique of Presser \textit{et al.} \cite{presser2004methods}, so participants who state they use this tool could be highlighted. Whilst false data is often discounted from datasets, we study PrivBrowse responses to evaluate respondents' perceptions of privacy-protective technologies.

\section{Results and Discussion}
\label{sec:four}

\subsection{Participants}

We collected a total of 112 physical responses from across the UK, with 94\% of the participants identifying themselves as UK residents. Our gender ratio was reasonably balanced, 57\% female and 43\% male, and at least 9\% of participants came from each age bracket. More than 10\% also came from each education level other than PhD, suggesting a varied cross-section of the British public. Our survey greatly benefits from being physically-conducted, as online polls can reduce participation from certain demographic groups \cite{evans2005value}. 71\% reported to use computers as part of their job, highlighting how ubiquitous technology has become in the UK.

\subsection{Analytical Techniques}

We first used Cronbach's alpha ($\alpha$) to measure our scale reliability, with coefficients in excess of 0.7 considered to represent internal consistency. Fortunately, our $\alpha$ = 0.81 and therefore fulfilled this requirement. \newpage

We utilised a number of techniques to study correlation and differentiate between datasets. We used Spearman's Rank Correlation Coefficient ($\rho$) to measure two-variable linear correlation from 1 for total positive correlation to 0 for no correlation to -1 for total negative correlation. This technique is used for ordinal data, which we receive from ranking our Likert scale responses numerically. We studied the \textit{p}-value to assess the probability the result is due to chance, with a \textit{p} \textless \xspace 0.05 representing statistical significance.

To determine whether two datasets were significantly different from each other, we used the Mann-Whitney-Wilcoxon test for our ordinal opinion data and the two-tailed Student's t-test otherwise. We again required \textit{p} \textless \xspace 0.05 to indicate differences were probabilistically not due to chance.

In investigating our research questions (RQ), we studied five variables: the mean amount of \textit{data disclosed} (0--3), the mean \textit{privacy opinions} (1--5), the mean \textit{online privacy opinions} (1--5), the mean privacy \textit{self-evaluations} (1--5), and the mean self-reported \textit{action scores} (1--4). This \textit{action score} ranged from most to least private, with ``Unsure'' and ``N/A'' answers deemed to represent a neutral response. In each case, the lower the value of the variable, the more `private' a participant was considered to be. For the remainder of the paper, we use $\bar{x}$ to denote mean values, $\rho$ for correlation coefficients and include \textit{p}-values when differences are statistically significant.

\subsection{RQ1: Privacy Opinions and Protective Technologies}

In answering our first research question, we analysed respondents' opinions and actions.
We found participants overwhelmingly claim to be concerned about their privacy. 92\% reported they found privacy important to them, 69\% strongly agreeing with the statement compared to 2\% in disagreement. Similarly, 93\% agreed \textit{online} privacy was important, with less than 3\% disagreeing with the statement. Participants were more reserved over self-evaluations of their own privacy, but 81\% still agreed that they acted privately. We now compare these opinions with reported actions to investigate the Privacy Paradox.

\begin{table}[ht]
\scriptsize
\setlength{\tabcolsep}{.425em}
\begin{tabular}{lrrrrrr} 
\toprule
Action & Always & Often & Rarely & Never & Unsure & N/A \\ 
\midrule
\pbox{18cm}{How often do you clean your Internet browser's  \\history?} 
 & 15.2
 & 31.3
 & 25.9
 & 23.2
 & 1.8
 & 2.7
 \\[2.5ex]
\pbox{18cm}{How often do you use Internet browser plug-ins/\\extensions to protect your privacy?} 
 & 15.2
 & 21.4
 & 12.5
 & 42.9
 & 5.4
 & 2.7
 \\[2.5ex] 
\pbox{18cm}{How often do you encrypt data on your\\ computer?} 
 & 6.3
 & 8.9
 & 15.2
 & 66.1
 & 0.9
 & 2.7
 \\[2.5ex] 
\pbox{18cm}{How often do you store unencrypted data\\ within a cloud provider such as Dropbox?} 
 & 17.9
 & 16.1
 & 13.4
 & 44.5
 & 1.8
 & 5.4
 \\[2.5ex] 
\pbox{18cm}{How often do you share public posts on social \\networking sites such as Facebook?} 
 & 17.9
 & 18.8
 & 17.0
 & 25.9
 & 0.9
 & 19.6
 \\[2.5ex] 
\pbox{18cm}{How often do you share your location on social \\networking sites?} 
 & 3.6
 & 10.7
 & 29.5
 & 36.6
 & 0
 & 19.6
 \\[2.5ex] 
\pbox{18cm}{How often do you use Tor for your web \\browsing?} 
 & 1.8
 & 2.7
 & 7.1
 & 75.9
 & 7.1
 & 5.4
 \\[2.5ex] 
\pbox{18cm}{How often do you use PrivBrowse for your web \\browsing?} 
 & 0
 & 6.3
 & 1.8
 & 82.1
 & 5.4
 & 4.4
 \\[2.5ex] 
\pbox{18cm}{How often do you use encryption tools for your \\emails?} 
 & 7.1
 & 8.9
 & 4.5
 & 72.3
 & 3.6
 & 3.6
 \\[2.5ex] 
\pbox{18cm}{How often do you read the terms and \\conditions on websites you use?}
 & 14.3
 & 18.8
 & 22.3
 & 40.2
 & 1.8
 & 2.7
 \\[2.5ex] 
\pbox{18cm}{How often do you check permissions before \\installing smartphone apps?}
 & 19.6
 & 11.6
 & 15.2
 & 30.4
 & 0.9
 & 22.3
 \\ 
\bottomrule
\end{tabular}
\normalsize
\centering
\caption{Self-reported action frequencies (\%)}
\label{table:actions}
\end{table}

Whilst the complete list of action results are shown below in Table \ref{table:actions}, we focus on privacy-protective technology use, looking to address our first research question. Not even half of the respondents claimed to clean their web browser history frequently, which is worrying considering less than 2\% were unsure of their answer. With the retention of browsing histories and cookies potentially enabling online tracking, this could place a large number of individuals at risk. 42\% of participants had not used protective browser plug-ins, despite the popularity of ad-blocking software. Data encryption is applied rarely, with only 6\% of respondents always performing the task. This comes in contrast to the 66\% who admitted to never encrypting at all. Less than 5\% of participants claimed to use Tor\footnote{https://www.torproject.org/} frequently, with 75\% having never used the software. Anecdotally, one respondent reported that the technology was used by criminals, and this stigma might explain its lack of common usage. Despite efforts to increase PGP\footnote{https://www.gnupg.org/} usability, still 72\% of individuals have never encrypted their emails. Only 7\% claimed to use email encryption consistently, and assuming our sample is somewhat representative, this suggests that the vast majority of UK public email is open to eavesdropping.

Despite the overwhelming concern claimed for privacy, with 92\% agreeing with its importance, individuals do not take actions to protect themselves. We found little correlation between \textit{privacy opinions} and \textit{action scores} (correlation coefficient \textit{$\rho$} = 0.02), reflecting that just because a participant claims to value privacy, it does not mean they act privately. \textit{Online privacy opinions} also had little correlation with  \textit{action scores} (\textit{$\rho$} = -0.03), suggesting the existence of the Privacy Paradox. Interestingly, we found no correlation between \textit{self-evaluations} and \textit{action scores} (\textit{$\rho$} = 0.06), which could indicate poor privacy self-awareness. Potential reasons for these disparities include a lack of privacy awareness, the inconvenience of using protective tools, or simply because individuals might not associate certain actions with acting privately.

Remarkably, 8\% of participants claimed to have used the PrivBrowse tool, despite it being fictitious. We found these individuals cared less about both privacy and online privacy than average but evaluated themselves similarly private. The group might have responded in this manner through a simple misconception or due to social desirability bias. \newpage

\subsection{RQ2: Demographic Comparisons}

For our second research question, we investigated whether younger adults care less about privacy, or are better able to protect themselves. We also studied how different genders regard the topic, and whether those better educated in privacy matters act more privately.

Whilst several age groups significantly differed in \textit{action scores}, these did not follow a clear pattern based on youth or age. Similarly, although we found several significant differences for \textit{online privacy opinions} and \textit{self-evaluations}, these did not follow a consistent trend. To investigate a direct distinction, we split respondents into 18--45 and 46 and above groups. Those less than 46 would have been born at least in 1970 and likely experienced the personal computer revolution of the 80s onwards. Although not significantly, we found older individuals acted more privately, disclosed less information, and rated themselves as more private. In contrast, younger adults reported to care more about both privacy and online privacy. We went on to examine whether younger respondents used privacy-protective technologies more frequently, and this was found to be true. Although the \textit{action scores} did not differ significantly, those under 46 years of age reported more frequent usage of all five technologies (browser history erasure, browser privacy plug-ins, data encryption, Tor and email encryption). This could dispel the myth that age correlates with privacy concern, as although older individuals acted more privately, those less than 46 use technologies more frequently to protect themselves.

We continued by studying whether gender, education or computer usage affected privacy perceptions. Regarding gender, we found women to be significantly more confident than men ($\bar{x}$ = 1.7 vs $\bar{x}$ = 2.1, \textit{p} = 0.01) through their \textit{self-evaluations}. Curiously, this did not translate into lower female \textit{action scores}, with men faring better than women. This appears to support earlier work by Sheehan \cite{sheehan1999investigation} which found that men are more likely to act to protect their privacy, represented by our male respondents also disclosing less data. Whilst significant differences were not seen through education level, those with Master's degrees acted most privately, disclosed the least information, and cared most about both privacy and online privacy. Those who used PCs during their employment had better \textit{action scores}, disclosed less data, cared more about privacy, and evaluated themselves as more private, though these differences were not significant in our study.

To further analyse whether knowledge affects privacy perceptions, after our public survey was completed we surveyed an additional 25 cybersecurity researchers. 80\% of this sample were male, 80\% were under 36, and all were UK residents with at least a Bachelor's degree. We found the researchers' \textit{action scores} were significantly more private than the general public ($\bar{x}$ = 2.4 vs $\bar{x}$ = 2.7, \textit{p} \textless \xspace 0.01), and they revealed fewer optional demographics. Education clearly has an effect: those with Master's degrees acted most privately in the public sample, with cybersecurity researchers disclosing even less data. This might suggest that privacy awareness campaigns and training sessions would be beneficial in encouraging the public to act more privately.

\subsection{RQ3: Data Disclosure Behaviour}

In answering our third research question, we investigated how data disclosure behaviour correlates against privacy opinions, actions, and self-evaluations. All three optional demographic queries possessed a ``Prefer Not to Say'' option and privacy should be a salient thought when completing a survey on the topic. 

In spite of this, 96\% of participants willingly disclosed their marital status, suggesting that individuals do not treat these details privately. Whilst 12\% of respondents revealed their Twitter handles, only 13\% explicitly chose not to. The remaining individuals claimed to not possess an account, and whilst this might be an anti-disclosure tactic, if more respondents used Twitter then the disclosure rate would likely increase. Since the handle enables direct contact with the user, it is still concerning that over one in ten participants would needlessly reveal this information. 56\% of respondents chose to reveal their employer, with this figure increasing to 84\% when including the unemployed and retired. On modal average, 72\% of respondents disclosed two items of data needlessly, with over 99\% revealing at least one piece of optional information. Assuming our sample is somewhat representative, the UK public are very willing to disclose their personal data.

\begin{figure}[ht]
  \centering
    \includegraphics[width=1.0\textwidth]{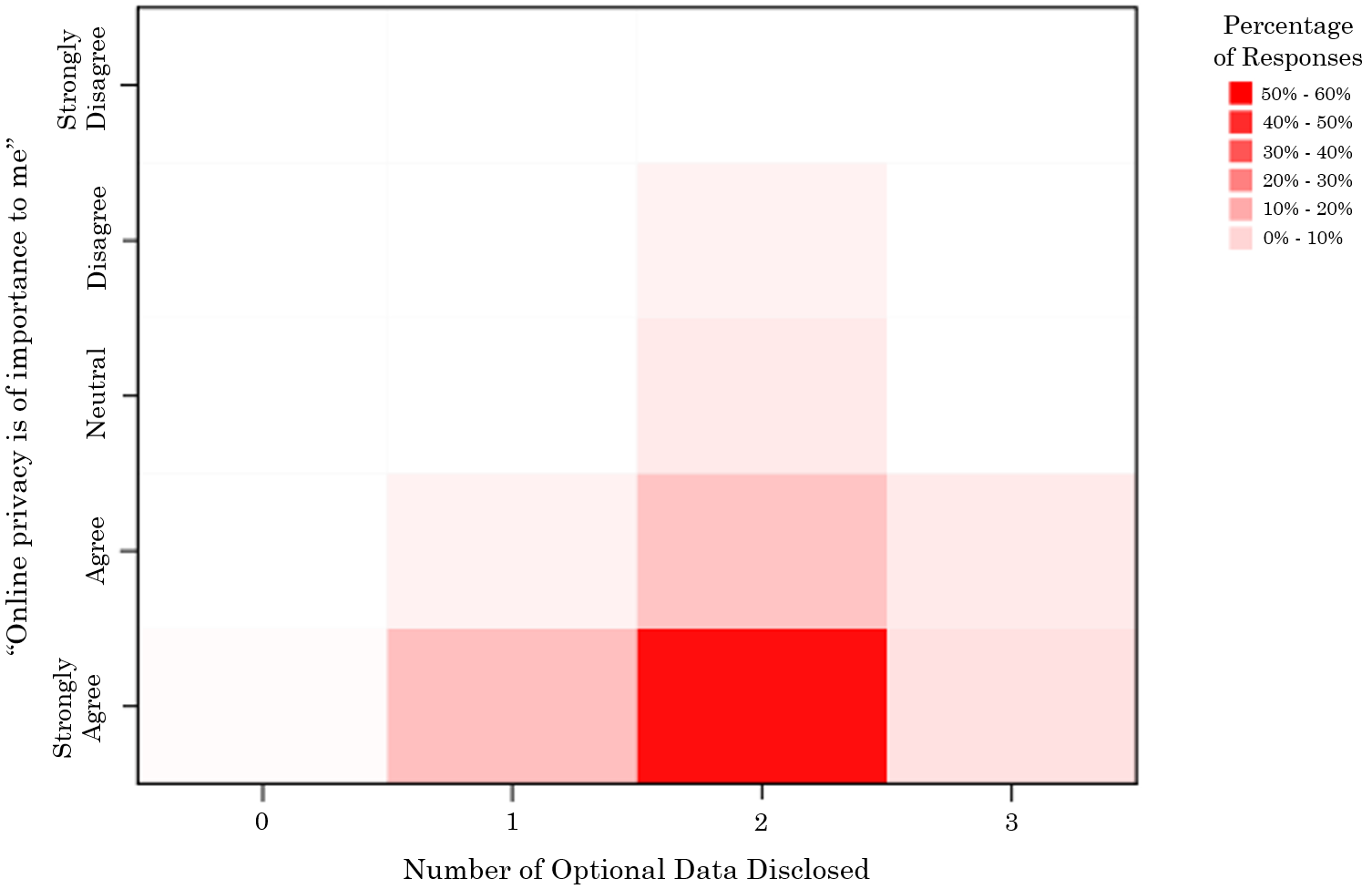}
    \caption{Rating of online privacy importance as compared to disclosure behaviour}
    \label{fig:heatmap}
\end{figure}

We finally studied the effect that opinions and self-evaluations have on data disclosure. We discovered that both \textit{privacy opinions} (correlation coefficient \textit{$\rho$} = 0.06) and \textit{online privacy opinions} (\textit{$\rho$} = 0.15) bore little correlation with the amount of \textit{data disclosed}, with the latter comparison shown above in Fig. \ref{fig:heatmap}. The heatmap suggests that regardless of how much an individual might profess to value online privacy, they are just as likely to divulge optional demographic information. \textit{Self-evaluations} (\textit{$\rho$} = 0.06), were also found to have a minimal effect on disclosure behaviour, suggesting individuals reveal data regardless of their beliefs. Of greatest concern, reported \textit{action scores} (\textit{$\rho$} = 0.05) did not correlate with \textit{data disclosed}, reflecting that even those who claim to act privately needlessly reveal their information. Here we illuminate another angle of the Privacy Paradox: not only do individuals report actions different to their opinions, they disclose information needlessly even whilst describing themselves as private.

\section{Conclusions}
\label{sec:five}

In this paper, we investigated the UK public's perceptions of online privacy, their use of privacy-protective tools, and how different demographic groups regard the topic. We found that whilst the vast majority of participants claim to value privacy, they do not act to keep their information safe. Although 93\% of respondents agreed with the importance of online privacy, fewer than half even cleaned their browser history regularly. This might be due to insufficient privacy education, apathy, or because respondents do not appreciate the risks they encounter in their use of online services \cite{Creese2012} \cite{Nurse2015xrds}. We saw that younger adults do not necessarily care less about online privacy, in fact they use protective technologies more frequently than their older counterparts. Education also played a strong role: those with Master's degrees acted most privately in the public sample, with cybersecurity researchers disclosing even less data. Finally, we discovered that information disclosure does not correlate with privacy opinions or reported actions. With 96\% of respondents divulging their marital status and more than half disclosing their employer with little need, the British public appear very willing to reveal their personal information. In total, over 99\% of individuals disclosed at least one piece of data needlessly, with over one in ten revealing their Twitter handle.  No correlation was found between participants' opinions and the actions they took, validating the existence of the Privacy Paradox. We develop the concept further to show that individuals disclose information needlessly even whilst describing themselves as private.

Whilst we hope our research will assist others in analysing data disclosure, we accept limitations to our work. Although our four selected cities provided variations in population size and location, we did not survey all areas of the UK. A more representative future work would cover cities in Scotland and Northern Ireland, canvas a larger number of sites, and include rural locations in addition to cities. Whilst we also do not consider 112 physically-surveyed respondents an insignificant sample, this figure could be increased through the future use of a mixed-mode survey including an online form.

With the Privacy Paradox receiving increased interest, there is much further work to conduct in this area. Firstly, a similar survey could be conducted online to examine whether individuals are more willing to disclose their data to a web form or to a physical researcher. Secondly, the effect of privacy salience could be studied against the level of information disclosure. We could conduct both a privacy survey and an unrelated poll to compare whether respondents disclose less data when actively considering privacy. Thirdly, we could conduct a European study to empirically compare privacy behaviour across different cultures. With privacy laws stricter in some states, such as Germany, we could study whether these respondents disclose less data than citizens of other countries.
\\

\textbf{Acknowledgments.} We would wish to thank the UK EPSRC who have funded this research through a PhD studentship in Cyber Security.

\vspace{-1em}

\bibliographystyle{splncs03}
\bibliography{bib}

\end{document}